\documentclass[fleqn,10pt]{wlscirep}
\usepackage[utf8]{inputenc}
\usepackage[T1]{fontenc}
\title{Non-Conventional Critical Behavior and Q-dependent Electron-Phonon Coupling Induced Phonon Softening in the CDW Superconductor LaPt$_2$Si$_2$}

\author[1,*]{E.~Nocerino}
\author[2]{U.~Stuhr}
\author[3,4]{I.~Sanlorenzo}

\author[5]{F.~Mazza}
\author[2]{D.~G.~Mazzone}
\author[6,7]{J.~Hellsvik}
\author[8]{S.~Hasegawa}
\author[8]{S.~Asai}
\author[8]{T.~Masuda}
\author[9]{S.~Itoh}
\author[10]{A.~Minelli}
\author[11]{Z.~Hossain}
\author[12]{A.~Thamizhavel}
\author[3]{K.~Lefmann}
\author[13]{Y.~Sassa}
\author[1,$\dagger$]{M.~M{\aa}nsson}

\affil[1]{Department of Applied Physics, KTH Royal Institute of Technology, SE-106 91 Stockholm, Sweden}

\affil[2]{Laboratory for Neutron Scattering and Imaging, Paul Scherrer Institute, CH-5232 Villigen PSI, Switzerland}

\affil[3]{Nanoscience Center, Niels Bohr Institute, University of Copenhagen, Nørre Allé 59, DK-2100 Copenhagen O, Denmark}

\affil[4]{Department of Applied Science and Technology, Politecnico di Torino, corso Duca degli abruzzi 24
10129 Torino, Italy}

\affil[5]{Institute of Solid State Physics, Vienna University of Technology, Wiedner Hauptstraße 8–10, 1040 Vienna, Austria}

\affil[6]{PDC Center for High Performance Computing, KTH Royal Institute of Technology, SE-100 44 Stockholm, Sweden}

\affil[7]{Nordita, KTH Royal Institute of Technology and Stockholm University, Hannes Alfvéns väg 12, SE-106 91 Stockholm, Sweden}

\affil[8]{Neutron Science Laboratory, Institute for Solid State Physics, The University of Tokyo, Kashiwa, Chiba 277-8581, Japan}

\affil[9]{Institute of Materials Structure Science, High Energy Accelerator Research Organization, Ibaraki 305-0801, Japan}

\affil[10]{Inorganic Chemistry Laboratory, University of Oxford, Oxford OX1 3QR, United Kingdom}
\affil[11]{Department of Physics, Indian Institute of Technology, Kanpur 208016, India}
\affil[12]{DCMPMS, Tata Institute of Fundamental Research, Mumbai 400005, India}

\affil[13]{Department of Physics, Chalmers University of Technology, SE-412 96 Göteborg, Sweden}

\affil[*]{nocerino@kth.se}

\affil[$\dagger$]{condmat@kth.se}

\begin{abstract}
This paper reports the first experimental observation of phonons and their softening on single crystalline LaPt$_2$Si$_2$ via inelastic neutron scattering. From the temperature dependence of the phonon frequency in close proximity to the charge-density wave (CDW) $q$-vector, we obtain a CDW transition temperature of T$_{CDW}$ = 230 K and a critical exponent $\beta$ = 0.28 $\pm$ 0.03. This value is suggestive of a non-conventional critical behavior for the CDW phase transition in LaPt$_2$Si$_2$, compatible with a scenario of CDW discommensuration (DC). The DC would be caused by the existence of two CDWs in this material, propagating separately in the non equivalent (Si1–Pt2-Si1) and (Pt1–Si2-Pt1) layers respectively, with transition temperatures T$_{CDW-1}$ = 230 K and T$_{CDW-2}$ = 110 K. A strong $q$-dependence of the electron-phonon coupling has been identified as the driving mechanism for the CDW transition at T$_{CDW-1}$ = 230 K while a CDW with 3-dimensional character, and Fermi surface quasi-nesting as a driving mechanism, is suggested for the transition at T$_{CDW-2}$ = 110 K. Our results clarify some aspects of the CDW transition in LaPt$_2$Si$_2$, which have been so far misinterpreted by both theoretical predictions and experimental observations, and give direct insight into its actual temperature dependence.
\end{abstract}
\begin{document}

\flushbottom
\maketitle
%
%
\thispagestyle{empty}


\section*{Introduction}

\begin{figure}[ht]
  \begin{center}
    \includegraphics[scale=0.4]{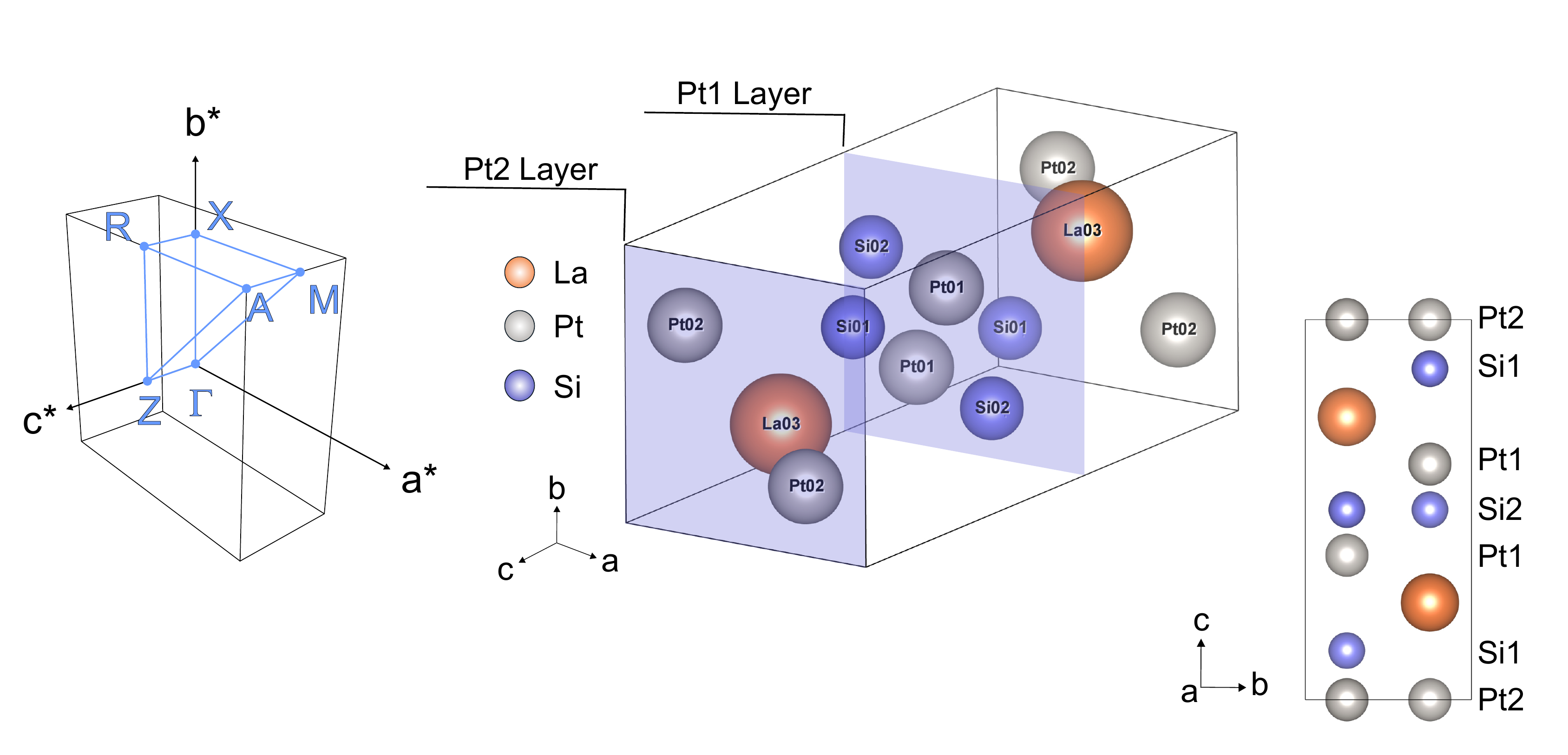}
  \end{center}
  \caption{Brillouin zone of the tetragonal $P4/nmm$ symmetry group, with the related symmetry points and axes, along with the room temperature unit cell of LaPt$_2$Si$_2$. The direct and reciprocal cells are depicted with the same orientation of the axes. The room temperature unit cell oriented along the $a$-axis is also shown for clarity of display of the Pt-Si layers}
  \label{brillouin}
\end{figure}

The concept of charge-density waves (CDW) is associated with the periodic spatial modulation of electron density in metallic systems, where the symmetry of the otherwise highly uniform charge density is broken due to instabilities of the Fermi Surface \cite{gruner1988dynamics}. In CDW materials, structural modifications of the crystal lattice, such as super lattice distortions and modifications of the crystal symmetry, usually occur as a consequence of the charge displacement to lower the energy of the system. Introduced for the first time in the 1950's \cite{frohlich1954theory}, CDWs were more recently brought back to the headlines since evidence of incommensurate charge ordering phenomena were found in several unconventional superconductors \cite{da2014ubiquitous, wagner2008tuning, wang2013crossover}. For this reason, the interplay between CDW and superconductivity (SC) is believed to be a key factor in understanding the mechanism behind unconventional superconductivity, which is one of the major unresolved questions in modern condensed matter physics.
LaPt$_2$Si$_2$ is a layered Pt-based rare earth intermetallic material which exhibits strong interplay between CDW and SC \cite{kim2015mechanism,shen2020evolution} and, therefore, represents a very interesting study case in this framework. The room temperature crystal structure of LaPt$_2$Si$_2$ is a CaBe$_2$Ge$_2$-type tetragonal structure with space group $P4/nmm$, where two alternating (Si1-Pt2-Si1) and (Pt1-Si2-Pt1) layers, separated by lanthanum atoms, are stacked along the $c$-axis \cite{NocerinoXRD} (Fig. \ref{brillouin}).

According to theoretical calculations \cite{hase2013electronic, kim2015mechanism}, the Fermi surface of LaPt$_2$Si$_2$ is expected to be of two-dimensional nature with electron-like pockets around the M-point. These pockets are quasi-nested with a $q$-vector $q_{CDW}$ close to (1/3, 0, 0). The quasi-nesting is due to the predominant contribution of the Pt1 atom matrix element to the calculated projected susceptibility, which implies dominant role of the Pt1 atoms in the CDW transition and partial gapping of the Fermi surface at $q$-values close to the nesting vector. Phonon dispersion calculations were also performed to account for the role of electron-phonon coupling in the CDW transition \cite{kim2015mechanism}; phonon-softening instabilities leading to structural instabilities, mainly due to the Pt1 atoms, were predicted to occur in correspondence to $q_{CDW} = (1/3, 0, 0)$. An estimation of the electron-phonon coupling constant $\lambda_p$ is also provided, and found to be consistent with a moderately strong coupling ($\lambda_p$ = 0.73). It was concluded that the CDW in LaPt$_2$Si$_2$ is 2-dimensional in nature and likely originates from $q$-dependent electron-phonon coupling with quasi-nesting of the Fermi surface. The CDW was claimed to propagate in the Pt1 layer, within which it would co-exist with superconductivity \cite{kim2015mechanism}. 

Experimentally, indications of the occurrence of the CDW transition in this material were observed with several experimental techniques in both powder samples and single crystals \cite{gupta2016coexistence, kubo2014structural, nagano2013charge, falkowski2019structural, aoyama2018195pt}. From these investigations, it was established that the CDW transition temperature should be around 85 K for single crystalline LaPt$_2$Si$_2$ or 110 K for polycrystalline LaPt$_2$Si$_2$, in correspondence with sharp anomalies observed in magnetic susceptibility, resistivity and specific heat data, as well as in the temperature dependence of the Knight shift and NMR relaxation rate extracted from $^{195}$Pt-NMR measurements. The appearance of super-lattice satellite Bragg reflections in the diffraction patterns of both single crystal and powder LaPt$_2$Si$_2$ \cite{nagano2013charge, falkowski2019structural} with propagation vectors $q\prime1$ = (0.36, 0, 0) and $q\prime \prime1$ = (0, 0.36, 0), provided additional indication of the occurrence of a CDW state. In single crystals, the satellites appear around 175 K, have their maximum intensity at 85 K and disappear below 80 K. For this reason, the CDW transition temperature was associated with the temperature at which the satellites have their maximum intensity (i.e., $T_{CDW}$ = 85 K). Later on, a second set of satellites appearing around 85 K was identified, with a periodicity $q\prime2$ = (0.18, 0.18, 0.5), $q\prime \prime2$ = (0.18, -0.18, 0.5) that can be expressed as a linear combination of the $q1$ satellites propagation vectors: 
\begin{eqnarray}
 q\prime2 & = \frac{q\prime1 + q\prime \prime1 + (001)}{2},\\ 
 q\prime \prime2 & = \frac{q\prime1 - q\prime \prime1 + (001)}{2}.
\label{eq2}
\end{eqnarray}
This finding seems to indicate that multiple CDW transitions take place in LaPt$_2$Si$_2$ and casted some doubts on the previous results, as well as on the actual temperature dependence of the CDW transition \cite{falkowski2020multiple}. Indeed, the reported T$_{CDW}$ = 85 K seems to be associated to the $q2$ satellites, while the $q1$ are probably related to a higher T$_{CDW}$ (the authors of reference \cite{falkowski2020multiple} suggest T$_{CDW}$($q2$) = 85 K and T$_{CDW}$($q1$) = 175 K). Our very recent high-resolution synchrotron XRD study, carried out within our collaboration, clarified the complex temperature dependent structural evolution of LaPt$_2$Si$_2$ \cite{NocerinoXRD}, which was found to undergo a series of structural transitions. Namely, on cooling, tetragonal $P4/nmm$ $\rightarrow$ incommensurate tetragonal $\rightarrow$ orthorhombic $Pmmn$ $\rightarrow$ tetragonal $P4/nmm$. The onset of the charge ordering that leads to the CDW transition was found to be well above room temperature and the first CDW transition was identified at T1 = 230 K, in correspondence to the formation of the $q1$ satellites, while a second CDW transition was identified at T2 = 110 K, in correspondence to the formation of the higher order $q2$ satellites \cite{NocerinoXRD}. The T1 transition was not considered earlier because the previous results were determined in a too narrow temperature range or with in-house experimental equipment, which did not have the necessary resolution to clearly identify the weak anomalies in their signals at T1 . 

To unambiguously determine whether the observed phenomena are really due to a CDW transition, and what is its actual temperature dependence, a measurement of the phonon dispersion curves in LaPt$_2$Si$_2$ is needed. This enables to verify the presence of a phonon softening, which is predicted theoretically to occur above T$_{CDW}$ to assist the CDW instability. In this work we present a direct observation of the phonon softening and T$_{CDW}$ determination in LaPt$_2$Si$_2$, measured with inelastic neutron scattering (INS).

\begin{figure*}[ht]
  \begin{center}
    \includegraphics[scale=0.75]{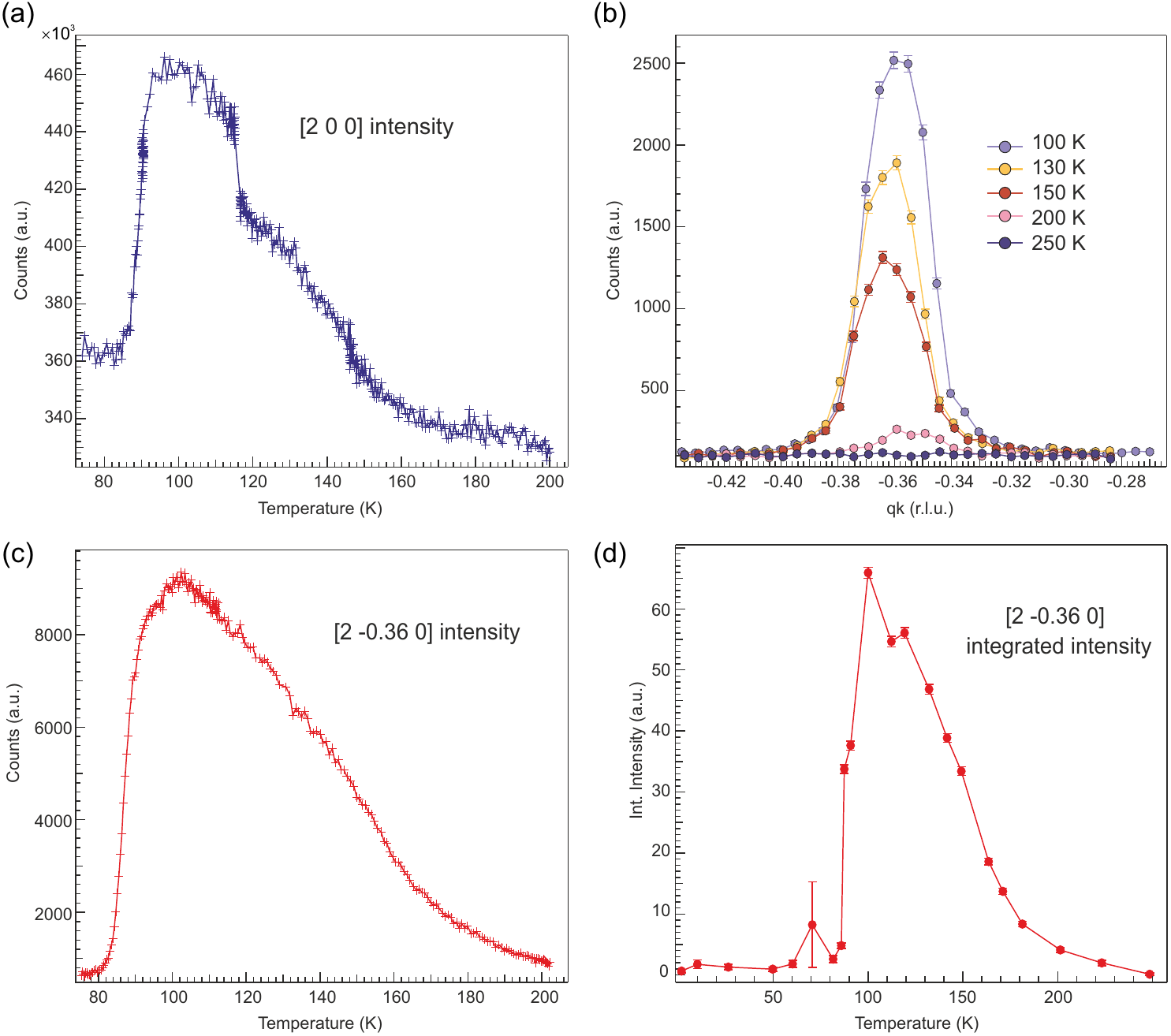}
  \end{center}
  \caption{a) Temperature dependence of the intensity of the [2 0 0] Bragg reflection. b) Temperature dependence of the $q1$ satellite [2 -0.36 0]. c) Temperature dependence of the intensity of the $q1$ satellite [2 -0.36 0]. d) Temperature dependence of the integrated intensity of the $q1$ satellite [2 -0.36 0].}
  \label{elastic_line}
\end{figure*}

\section*{Results}

The present study was conducted on two samples, a large single crystal with mass $\approx$ 4g and a smaller crystal with mass $\approx$ 1g. Due to its large mass, the big sample allowed the collection of INS scans with lower counting time with respect to the small sample, resulting in faster measurements with a better signal to noise ratio, and a consequent outline of the phonon dispersion relation with higher density of points. However, a twinning of about 8$^{\circ}$ in the large sample, with inequivalent contributions from the different twins, caused a systematic shift of the momentum transfer axis for the different energy transfer scans. The shift was estimated to be about 0.05 r.l.u. along $qk$, from calculations of the center of mass of the twin system and by direct comparison with the small, non twinned, sample. The behavior of the two crystals is in very good agreement modulo a 0.05 r.l.u. shift of the $qk$ axis, therefore, from now on we will refer to the measurements performed on the small non-twinned sample, while the plots collected for the large twinned sample (displayed in the Methods section) will be used as a qualitative term of comparison to facilitate the observation of the critical behavior in LaPt$_2$Si$_2$. It should be noted that the twinning in the large crystal could not be detected with surface characterization methods (i.e. X-ray Laue diffraction).

\begin{figure*}[ht]
  \begin{center}
    \includegraphics[scale=0.92]{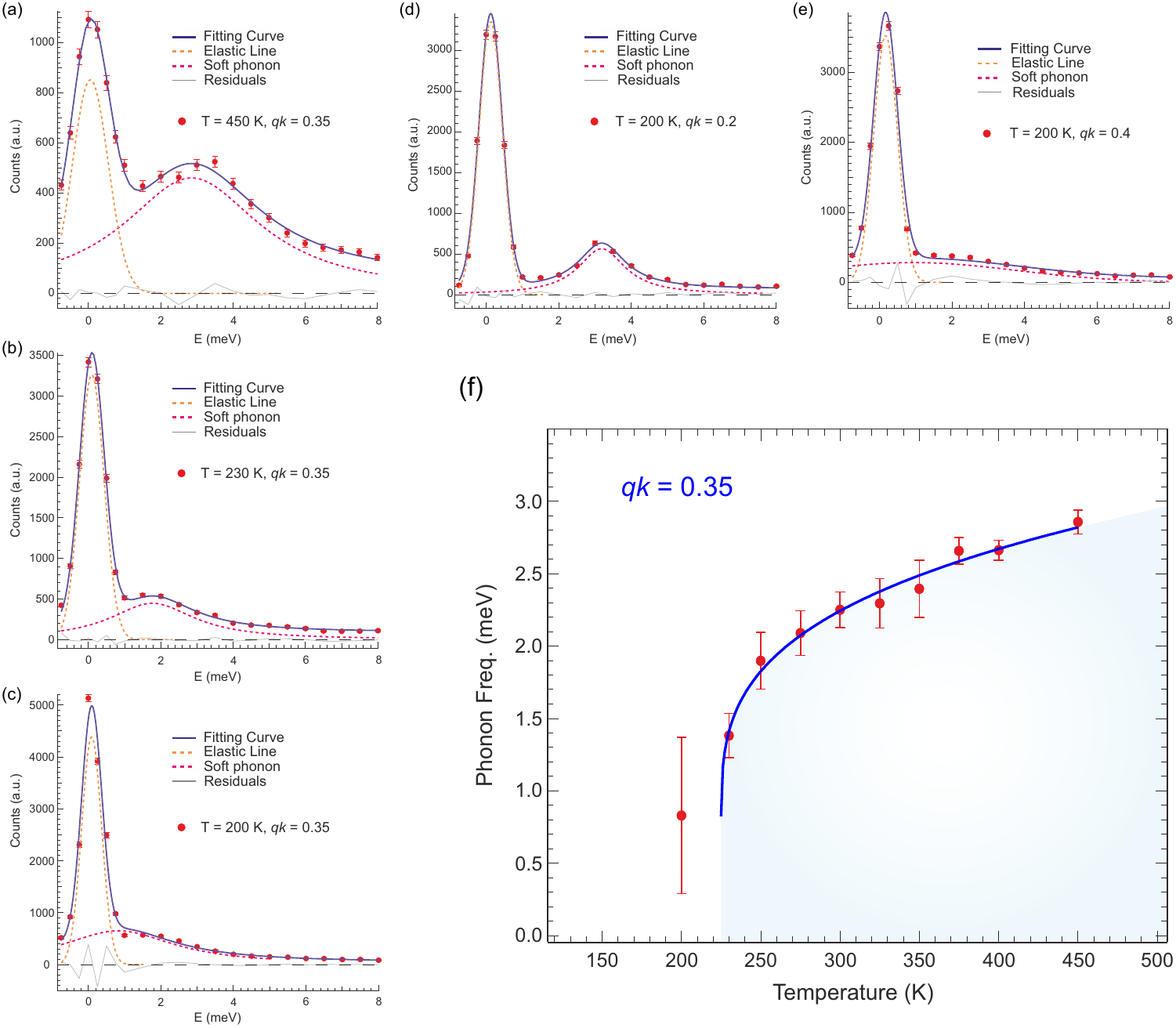}
  \end{center}
  \caption{(a-c) TAS-INS energy scans with model fits, taken at $qk$ = 0.35 for different temperatures showing the collapse of the phonon mode in the beginning of the critical region. (d-e) TAS-INS energy scans with model fits, taken at different $q$ points showing the $q$-dependence of the soft phonon mode at 200 K. (f) Temperature dependence of the phonon energy at qk = 0.35. The solid blue line is a fit to the power law $E_{tran} \propto (\frac{T}{T_c} - 1)^{\beta}$.}
  \label{ord_par}
\end{figure*}

A preliminary inspection of the zero energy transfer region of the spectrum (i.e., the elastic line), for alignment purposes on the Bragg reflection (2 0 0), gave us the opportunity to follow the temperature dependence of such peak, as well as of the $q1$ satellite (2 -0.36 0). Since, unlike X-rays, neutrons scatter with the atomic nuclei rather than with the surrounding electronic clouds, this measurement allows to investigate the temperature evolution of the lattice distortions without the charge ordering contribution. The outcome of this investigation can be therefore compared with the structural characterization suggested in reference \cite{NocerinoXRD}, to discriminate the effects of charge ordering and structural distortions on the diffraction data. Figure \ref{elastic_line}(a) displays the temperature dependent (2 0 0) structural Bragg peak intensity between 75 K and 200 K, whose behavior is in very good agreement with the findings of ref. \cite{NocerinoXRD}. In fact, on cooling, a smooth increase of the peak intensity is observed in correspondence to the incommensurate distortion in the $ab$-plane. Then two clear first order modulations of the peak intensity occur, one at $\sim$ 115 K, in proximity of the T2 = 110 K structural transition, and one at $\sim$ 90 K, in proximity of the strong fluctuations in the $b-a$ difference, that precede the structural transition at T3 = 60 K (here out of the investigated T-range) \cite{NocerinoXRD}. Figure \ref{elastic_line}(b-d) show the temperature dependence of the intensity and of the integrated intensity of the (2 -0.36 0) satellite. Also in this case, these parameters follow the trend outlined in reference \cite{NocerinoXRD}. Here it can be seen that the intensity of the $q1$ satellite is completely suppressed at 250 K, suggesting that the diffuse scattering observed in the XRD above this temperature is purely due to charge ordering.

Concerning the inelastic part of the spectrum, by following the transverse acoustic phonon branch of the (2 0 0) reflection along the $\Gamma - X$ direction (Fig. \ref{brillouin}), the phonon dispersion relation for different temperature points was constructed with constant $q$ and constant energy transfer scans. In order to follow the dynamical behavior of the phonon frequency as a function of the wavevector $qk$ for different temperatures, the inelastic part of the $q$ and energy scans were fitted to Lorentzian lineshapes, which are conventionally used to approximate phonon lineshapes to damped harmonic oscillators \cite{faak1997phonon}, while the elastic part was modeled with Gaussian lineshapes. Figure \ref{ord_par} (a-c) display the temperature dependence of the soft phonon mode at the $q$ point (2 0.35 0), in close proximity to $q_{CDW}$ = 0.36 (r.l.u.). Figure \ref{ord_par} (d-e) shows the $q$-dependence of the soft phonon mode below the CDW transition at $T=200$~K, outside and inside the critically damped region. The phonon is well shaped as a Lorentzian peak outside the critical region for $qk$ = 0.2 (r.l.u.). At $q_{CDW}$ the phonon appears very broad already at $T=450$~K [Fig. \ref{ord_par}(a)] and, at $T=200$~K additional broadening and shift in frequency result in a collapse of the mode on the elastic line [Fig. \ref{ord_par}(c)]. The clear dramatic change in the phonon position and broadening made it unnecessary for us to deconvolute the instrumental resolution function from the Lorentzian fit function, since the phonon linewidths were much larger than the instrumental resolution at EIGER ($<$ 1 meV in the energy transfer region of interest \cite{stuhr2017thermal}).

\begin{figure}[ht]
  \begin{center}
    \includegraphics[scale=0.42]{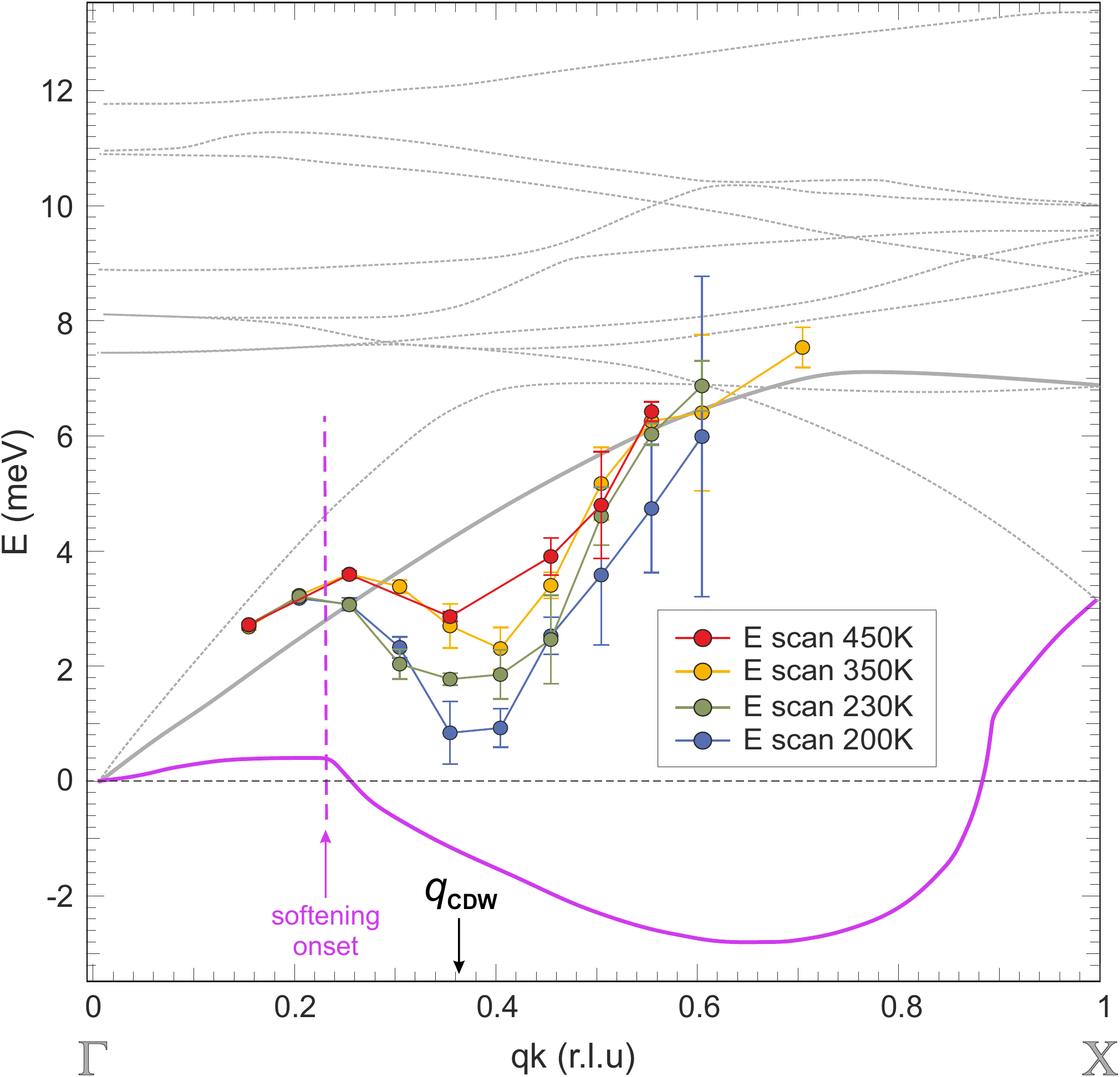}
  \end{center}
  \caption{TAS-INS phonon dispersion curves for different temperature points. The scatter plot (solid symbols) shows the experimental data, the grey lines is the lower energy part of the phonon dispersion calculated in reference \cite{kim2015mechanism} and the magenta line is the calculated soft mode.}
  \label{phonon_dataVStheory_unshifted}
\end{figure}

Additional temperature points were collected to observe the temperature dependence of the value of the phonon frequency at the $q$ point (2 0.35 0) [Fig. \ref{ord_par}(f)], which follows a power law of the type $E \propto (\frac{T}{T_c} - 1)^{\beta}$. Fitting the temperature dependence of the phonon frequency to the power law, the critical exponent was found to be $\beta = 0.29 \pm 0.21$, which implies critical behavior not well described by a mean-field theory approach, foreseeing an exponent $\beta = \frac{1}{2}$. This value is consistent with the value $\beta = 0.28 \pm 0.03$ extracted from the data on the twinned sample in $qk$ = 0.4 (r.l.u.) [see Fig. \ref{ord_par_twin}(f)], which could be estimated with higher confidence due to the larger number of experimental points in proximity of the phase transition. The fit also provides the transition temperature to the CDW state, $T_{CDW}$ = (225 $\pm$ 3) K, while the value estimated from the twinned sample is $T_{CDW}$ = (237 $\pm$ 4) K. These temperatures are reasonably close to each other 
and they are both in very good agreement with the transition temperature $T_1$ = 230 K suggested in reference \cite{NocerinoXRD}, as well as with the temperature dependence of the (2 -0.36 0) satellite displayed in Fig.~\ref{elastic_line} .
Figure \ref{phonon_dataVStheory_unshifted} displays the phonon dispersion curves constructed through the Lorentzian fitting for different temperature points. The experimental data are overlapped with the phonon dispersion previously calculated \cite{kim2015mechanism}, showing a very good agreement with the theoretical curves. Indeed, the measured onset of the softening is aligned with its counterpart in the calculated soft phonon mode. The $q$-dependent damping of the phonon mode is strongly pronounced already at 450 K, with a valley in the value of the phonon frequency around the $q$ point corresponding to $q_{CDW}$ (Fig. \ref{phonon_dataVStheory_unshifted}). The extent of the valley can be better appreciated looking at the phonon dispersion curves for the twinned sample (see Fig.~\ref{ord_par_twin}(g) in the Methods section). Here, the dispersion curve at T = 470 K shows a pronounced anomaly in a broad $q$ region, from qk = 0.3 (r.l.u.) to qk = 0.6 (r.l.u.). Higher temperatures were not accessible with our setup, however the anomaly in the phonon dispersion observed at T = 470 K implies that the onset of the CDW transition in this material occurs well above room temperature, confirming the indications of the synchrotron XRD results \cite{NocerinoXRD}.

\begin{figure*}[ht]
  \begin{center}
    \includegraphics[scale=1.1]{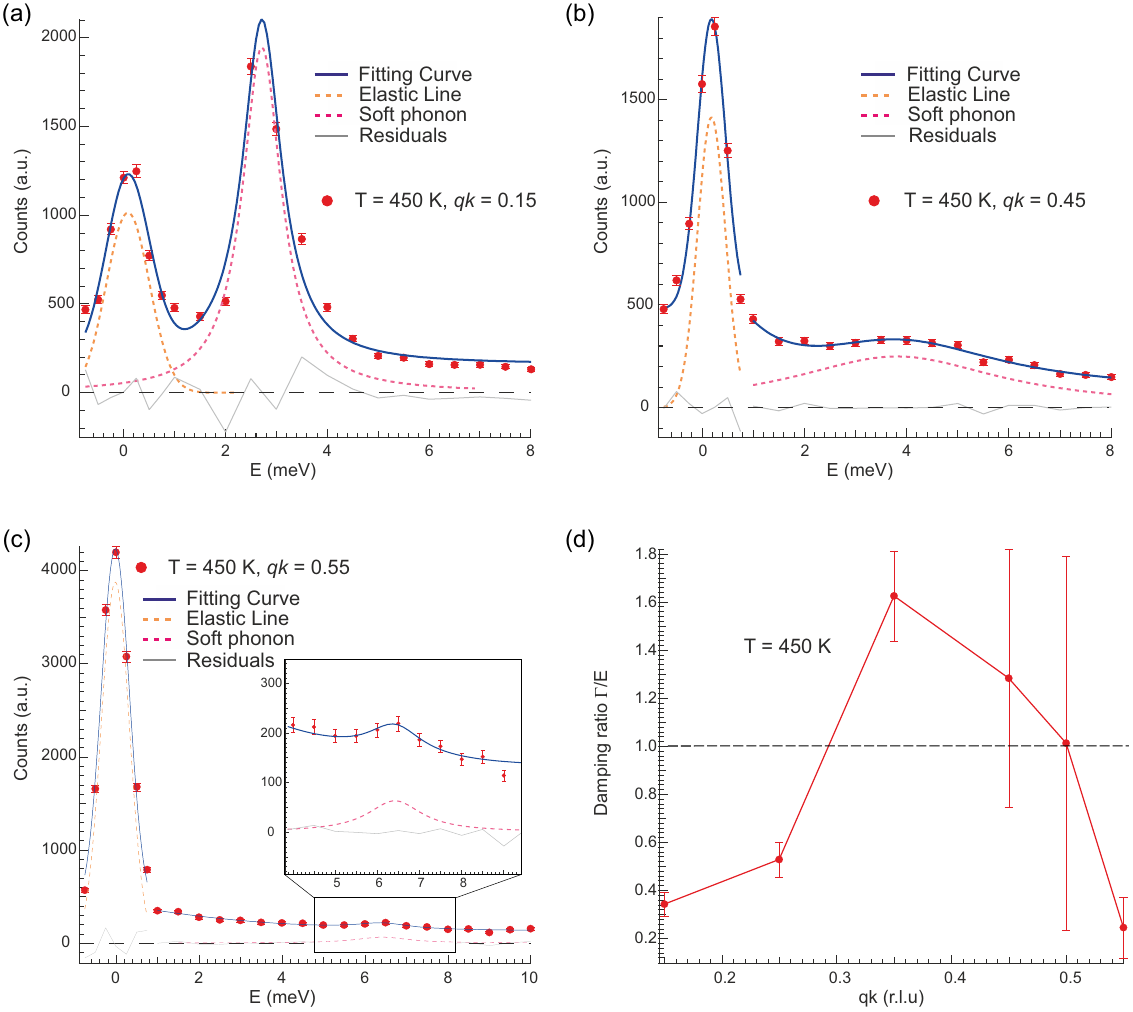}
  \end{center}
  \caption{(a-c) $q$-dependence of the soft phonon mode at $T=450$~K well outside the critical region. (d) $q$-dependence of the damping ratio for the phonon mode at $T=450$~K. The red dashed line is a guide to the eye and the black dashed horizontal line marks the overdamping condition threshold. Data taken by TAS-INS.}
  \label{damping}
\end{figure*}

On cooling from 450 K, the anomaly is more and more pronounced until the collapse of the phonon. Below 200 K the value for the phonon frequency in the critical region increases slowly on cooling (see. Fig. \ref{ord_par_twin} f in the Methods section). In $q$ regions above $q$ $>$ 0.4 (r.l.u.) the phonon frequency increases with increasing $q$, recapturing the regular transverse phonon profile. However, the shape of the phonon is not completely recovered and a very broad lineshape persists also outside of the critical region together with a dramatic intensity drop (Fig. \ref{damping}). One of the possible mechanisms that might justify such a damping of the phonon could be anharmonicity, as a direct consequence of the structural instability induced by the CDW transition.

With the observed renormalization of the phonon dispersion, over a wide range of wavevectors, and the broad critically damped region, which extends for an interval of wavevectors $\Delta$qk = 0.075 (r.l.u.), LaPt$_2$Si$_2$ exhibits a behavior that cannot be explained within the realm of the Peierls' picture, which foresees a sharp cusp-like deep in the phonon dispersion (Kohn anomaly) \cite{kohn1959image} in correspondence to the Fermi surface nesting (FSN) vector.
A similar phenomenology, with the broad Kohn-like anomaly in the phonon dispersion, occurs in the prototypical 2-dimensional CDW material 2$H$-NbSe$_2$ \cite{weber2011extended}. Here, the phonon softening instability was motivated uniquely by a strong $q$-dependence of the electron phonon coupling, since no indication of FSN was found. Indeed, looking at the temperature dependence of the resistivity in LaPt$_2$Si$_2$ \cite{NocerinoXRD} and 2$H$-NbSe$_2$ \cite{naito1982electrical}, no sharp metal to insulator transition (typical of the FSN-driven CDW) can be observed at their respective CDW transition temperatures (namely 230 K and 33.5 K respectively). Their resistivity only shows a deviation from linearity, with a downwards bending, which can be explained with CDW induced small anisotropic energy gaps which hardly affect the transport properties in these materials \cite{borisenko2009two}.
However, the two systems present two major differences between their dispersion relations. The first is that 2$H$-NbSe$_2$ does not present anharmonicity and the shape of the phonon is recovered at higher $q$ after the anomaly. The second difference is that, unlike 2$H$-NbSe$_2$, LaPt$_2$Si$_2$ shows an abrupt increase in the resistivity and a sharp drop in the magnetic susceptibility temperature dependencies occur at lower temperatures, in correspondence to the appearance of the $q2$ satellites and the first order structural transition $IC tetragonal$ $\rightarrow$ $orthorhombic$ $Pmmn$ at T2 = 110 K \cite{NocerinoXRD}. These sharp anomalies are evocative of a Peierls' transition, however they do not induce a full metal to insulator transition, because the system is still metallic below the transition. Therefore, they are compatible with partial anisotropic gaps opening at the Fermi surface, with a phenomenology similar to the quasi-1-dimensional system NbSe$_3$ \cite{hodeau1978charge,monceau1976electric}, in which two anomalies in the resistivity temperature dependence were observed. Here, the anomalies were explained with partial gaps opening at the Fermi surface, which were identified with an imperfect Fermi surface nesting. Nevertheless, unlike NbSe$_3$ in which the transitions in the electronic states are of second order, in LaPt$_2$Si$_2$ such transitions exhibit a hysteresis, which is signature of a first order transition. Since in regular 1D and 2D systems the CDW transition is usually of second order \cite{zhu2017misconceptions}, in reference \cite{NocerinoXRD} it is suggested that the abrupt changes in the resistivity and magnetic susceptibility in LaPt$_2$Si$_2$ are associated to the structural transition-induced changes in its electronic states. 

A plausible origin for the transition at T2 = 110 K with propagation vectors $q\prime2$ = (0.18, 0.18, 0.5), $q\prime \prime2$ = (0.18, -0.18, 0.5) is that a second CDW is established in LaPt$_2$Si$_2$, which has a different nature with respect to the one found at T$_{CDW}$ = 230 K. Indeed, the first order transitions in the electronic states accompanied by a first order restructuring of the atomic lattice with long-range ordering in the out-of-plane $c-$direction and strong inter-planar coupling \cite{NocerinoXRD}, might be consistent with a CDW transition with a 3-dimensional character and a quasi-nesting feature of the Fermi surface, which is indeed foreseen in the theoretical work of Kim et al. \cite{kim2015mechanism}. 3-dimensional CDW systems similar to LaPt$_2$Si$_2$, with 2D planar structures containing $5d$ transition metal ions with partially filled electronic levels such as IrTe$_2$ \cite{yang2012charge}, exhibit an analogous behavior of their transport properties within the CDW transition. Such systems belong to a class of structurally quasi 1D and quasi 2D materials, in which strong inter-chain and inter-planar couplings as well as spin-orbit coupling are believed to be related to the CDW transition \cite{becker1999strongly}. For this reason they are called "strong coupling CDW" materials or "3D-CDW" materials. In reference \cite{NocerinoXRD} LaPt$_2$Si$_2$ was associated with the 3D-CDW compound 2H-TaSe$_2$, due to the striking similarity between the behaviors of their structural evolution across the CDW phase transition \cite{moncton1975study}. Since discommensuration of the CDW was directly observed in 2H-TaSe$_2$ \cite{chen1981direct}, by virtue of their similarities, it was suggested that the CDW in LaPt$_2$Si$_2$ also undergoes discommensuration, with large commensurate domains separated by incommensurate narrow domain walls \cite{mcmillan1976theory}. Remarkably, the determination of the critical exponent in 2H-TaSe$_2$ \cite{brusdeylins1989determination} provided a value $\beta = \frac{1}{3}$, which implies a non conventional critical behavior for the CDW transition in 2H-TaSe$_2$ and is close, within the error bars, to the critical exponent estimated in this work for LaPt$_2$Si$_2$ $\beta = 0.28 \pm 0.03$. In qualitative agreement with 2H-TaSe$_2$, the non conventionality of the critical behavior of the CDW in LaPt$_2$Si$_2$ might be motivated by discommensuration, hereby endorsing the conjecture of reference \cite{NocerinoXRD}.
In LaPt$_2$Si$_2$ the 2 non-equivalent Pt1 and Pt2 sites give different contributions to the electronic density of states at the Fermi level, according to first principle calculations \cite{hase2013electronic,kim2015mechanism}. Additionally, experimental NMR measurements, acquired in a temperature range below 200 K, show that the Pt1 layer is the only responsible for the occurrence of the CDW state in LaPt$_2$Si$_2$ \cite{aoyama2018195pt}. In this regard, it is reasonable to believe that the proposed two CDW transitions at T$_{CDW-1}$ = 230 K and T$_{CDW-2}$ = 110 K have separate origins in the respective Pt planes, as also conjectured in reference \cite{NocerinoXRD}.
In brief, based on the arguments here in agreement with the published theoretical calculations \cite{kim2015mechanism} as well as with experimental evidences \cite{NocerinoXRD}, and in qualitative agreement with the phenomenology of analogous systems \cite{weber2011extended,naito1982electrical,yang2012charge,moncton1975study}, we suggest that the CDW formation in LaPt$_2$Si$_2$ is driven by two distinct mechanisms:

\begin{enumerate}
  \item a $q$-dependent electron-phonon coupling driven CDW transition, with a 2D character, which affects the Pt2 plane with transition temperature T$_{CDW-1}$ = 230 K,
  \item a Fermi surface quasi-nesting driven CDW transition, with a 3D character, which affects the Pt1 layer with T$_{CDW-2}$ = 110 K.
\end{enumerate}

This scenario foresees interaction between the two CDW mechanisms and is consistent with the suggested discommensuration scenario.

The softening of the phonon dispersion in LaPt$_2$Si$_2$ can be attributed to a strong wavevector dependence of the electron phonon coupling. As a side remark, it is interesting to notice that a $q$-dependent overdamping of the phonon is already in place at 450 K. Figure \ref{damping} displays the ratio between the linewidth of the phonon $\Gamma$, extracted from the Lorentzian fit, and the phonon frequency as a function of the wavevector for T = 450 K. From the plot it is evident that the overdamping condition $\frac{\Gamma}{\hbar\omega}>1$ (as described in ref.~\cite{weber2011electron}) is satisfied in a broad range of wavevectors already at this high temperature.
Since the electron-phonon coupling is believed to be the only mechanism involved in the first CDW transition at T$_{CDW-1}$ = 230 K observed through the phonon softening, it can be considered also the only responsible for such a dramatic enhancement of the phonon linewidth. Such a clear influence of the electron phonon coupling in LaPt$_2$Si$_2$ at a temperature as high as 450 K might have significant implications on the mechanism of superconductivity in LaPt$_2$Si$_2$, which is currently debated \cite{das2018multigap,nie2021nodeless}. Superconductivity in this material is not the subject of the current investigation, nevertheless doping dependent studies aimed at the suppression of the CDW transition might clarify this aspect.

An alternative explanation for the origin of the phonon broadening at high temperature might be the strong interaction between the phonons and the CDW. Indeed, phonon scattering due to charge fluctuations, which are inevitably present as precursors of the CDW, might be in place already at T = 450 K. Since these fluctuations disturb the periodicity of the lattice, they could act as scattering centers. This might be responsible for a very short lifetime of the phonons, evident from the observed broad linewidths \cite{gold2018acoustic}. 

\begin{figure*}[ht]
  \begin{center}
    \includegraphics[scale=0.85]{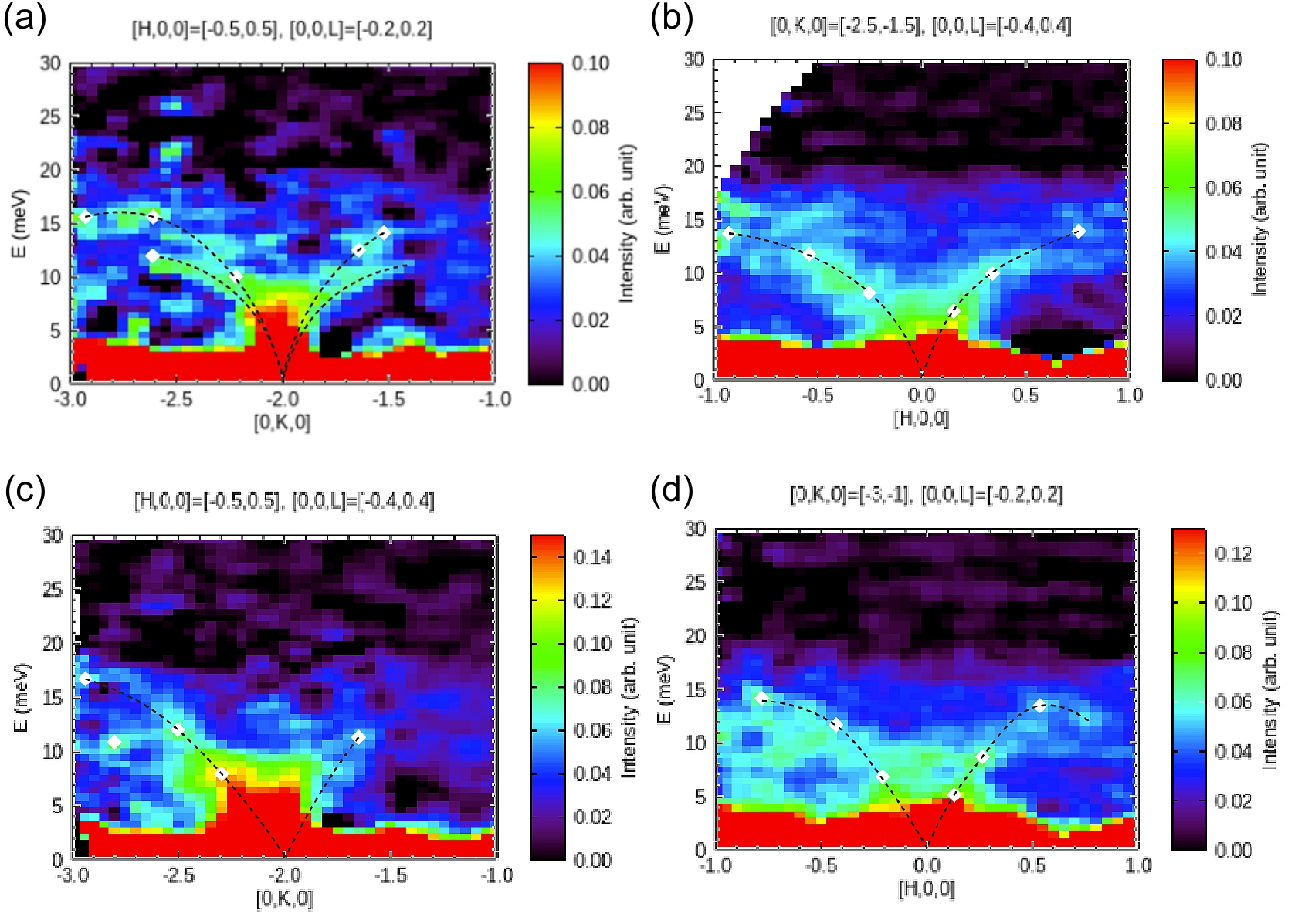}
  \end{center}
  \caption{Overview of the phonon modes for the Bragg point (0 -2 0). (a) Longitudinal phonon mode at $T=85$~K. (b) Transverse phonon mode at $T=85$~K. (c) Longitudinal phonon mode at $T=220$~K. (d) Transverse phonon mode at $T=220$~K. Data taken by ToF-INS}
  \label{overview_HRC}
\end{figure*}

This interpretation would also be consistent with the strong CDW-induced structural instability of the LaPt$_2$Si$_2$ lattice \cite{NocerinoXRD}.
An overview of the dispersion relation at 220 K and 85 K for the $q$ point (0 -2 0), provided by the TOF-INS HRC measurements, is shown in figure \ref{overview_HRC}. The dashed lines represent an attempt to determine the dispersion curves. However, the evident broadening and weak intensity of all the phonons in this materials makes the reliability of this approach questionable. In the figure we can see that the phonon dispersion profile appears very blurred, with overlapping of the acoustic modes and a relatively low intensity. Such a feature for the phonon spectra evidences the anharmonic character of the vibrational modes in this material \cite{wei2021phonon}.

Phonon anharmonicity was found to be related to ultralow thermal conductivity in several thermoelectric materials \cite{voneshen2013suppression,voneshen2017hopping,wakamura1990observation}. Under the light of this fact, the apparent phonon anharmonicity observed in the measured dispersion of single crystalline LaPt$_2$Si$_2$, is consistent with the value of thermal conductivity measured on a polycrystal sample of LaPt$_2$Si$_2$, $\kappa$ = 2.7 $\frac{W}{mK}$, with an estimated lattice-only thermal conductivity $\kappa_L \approx 0.1 \frac{W}{mK}$ at room temperature \cite{gupta2018thermal}. This value is remarkably low compared to the values of thermal conductivity commonly found in other electrically conducting materials (e.g. stainless steel has $\kappa$ = 15 $\frac{W}{mK}$). Due to its effect on $\kappa_L$, the phonon anharmonicity is considered to be a key factor in increasing the efficiency of thermoelectric materials, which are known to be extremely promising for efficient clean energy conversion \cite{bell2008cooling}. Therefore, due to its thermal transport properties and vibrational landscape, LaPt$_2$Si$_2$ seems to constitute also an interesting study case for the development of thermoelectric devices.

\section*{Discussion}

In this work we have observed the predicted phonon softening in the CDW superconductor LaPt$_2$Si$_2$. Following its temperature dependence, we could determine T$_{CDW}$ = 230 K. Indeed, the power law fit to the temperature dependence of the phonon frequency in the reciprocal lattice point $q$ = (2 0.35 0), in close proximity to $q_{CDW}$, confirmed the findings of reference \cite{NocerinoXRD}, posing the CDW transition temperature at T$_{CDW}$ = 230 K. 
From the same power law, the critical exponent of the phase transition was extracted as $\beta$ = 0.28, suggesting an unconventional critical behavior for the CDW in LaPt$_2$Si$_2$ and endorsing the conjecture that this material manifests CDW discommensuration.
We propose strong $q$-dependence of the electron-phonon coupling as the driving mechanism for the CDW transition at T$_{CDW-1}$ = 230 K in LaPt$_2$Si$_2$ and conjecture the possibility of a second CDW transition at T$_{CDW-2}$ = 110 K, related to the first one, with a 3-dimensional nature and Fermi surface quasi-nesting as a driving mechanism. The two CDWs should propagate separately in the non equivalent Pt2 and Pt1 layers respectively. While a strong hint of the the existence of the first mechanism is found in this work, additional studies would be necessary to prove also the second mechanism. Namely angle resolved photoemission spectroscopy (ARPES) for direct observation of the Fermi contour and electronic band structure, to confirm the Fermi surface nesting and probe the spin orbit coupling, as well as additional neutron spectroscopy studies, to identify eventual phonon softening associated to the second transition.

\section*{Methods}

LaPt$_2$Si$_2$ single crystals were prepared using arc melting of high purity La, Pt and Si at TIFR, Mumbai.
The phonon spectra were collected at the triple-axis inelastic neutron scattering (TAS-INS) instrument EIGER at the Paul Scherrer Institut (Switzerland) \cite{stuhr2017thermal} in the temperature range from 100 K to 470 K with a closed cycle cryo-furnace. The experiment at EIGER was carried out using a double focusing pyrolytic graphite monochromator and a horizontal focusing analyzer at a fixed energy E$_f$ = 14.68 meV. The room temperature unit cell of LaPt$_2$Si$_2$, refined from synchrotron x-ray diffraction data, was used for the mapping of the reciprocal space and for defining the momentum transfer \textbf{Q} = h\textbf{a*}+k\textbf{b*}+l\textbf{c*} which is expressed in reciprocal lattice units (r.l.u.) as (qh, qk, ql).
The crystal was aligned with the $hk0$-plane in the scattering plane. A low temperature overview of the phonon spectra was collected at the time-of-flight spectrometer (TOF-INS) HRC at JPARC (Japan) at the temperatures 3 K (here not shown), 85 K and 220 K with a Gifford McMahon (GM) bottom loading cryostat. The measurements in EIGER were performed in inverse geometry while in HRC in direct geometry. All images involving crystal structure were made with the VESTA software \cite{momma}, the TOF-INS data reduction was carried out with the software Hana, the $q$-E slices in the TOF-INS data were produced with the software Dave \cite{azuah2009dave}, the data plots were produced with the software IgorPro \cite{igor}.

Here we also show the data collected on the twinned sample. 

\begin{figure*}[ht]
  \begin{center}
    \includegraphics[scale=0.48]{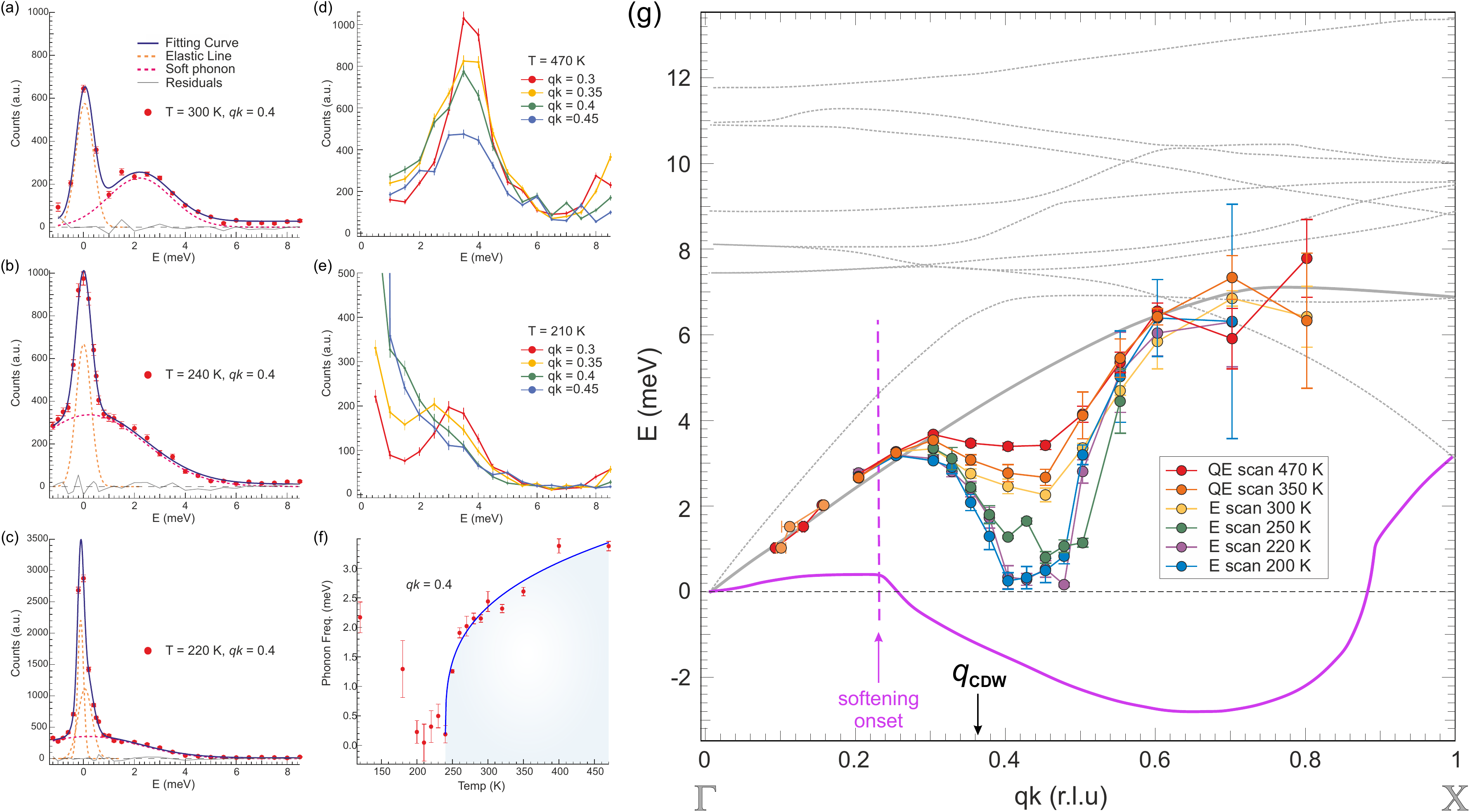}
  \end{center}
  \caption{\textbf{Data from Twinned Sample: }(a-c) TAS-INS energy scans with model fits, taken at qk = 0.4 for different temperatures showing the collapse of the phonon mode in the beginning of the critical region. (d-e) TAS-INS energy scans, taken at different $q$ points showing the $q$-dependence of the soft phonon mode at 470 K and 210 K, the solid lines are guides to the eye for clarity of display. (f) Temperature dependence of the phonon energy at qk = 0.45. The solid black line in (f) is a fit to the power law $E_{tran} \propto (\frac{T}{T_c} - 1)^{\beta}$. (g) TAS-INS phonon dispersion curves for different temperature points. The scatter plot (solid symbols) shows the experimental data collected on the twinned sample, the grey plot is the lower energy part of the phonon dispersion calculated in reference \cite{kim2015mechanism} and the magenta solid line is the calculated soft mode.}
  \label{ord_par_twin}
\end{figure*}

The energy scans at different $q$ points can be compared directly with the non twinned sample, modulo a shift of 0.05 (r.l.u.). The higher density of points in the dispersion relation and in the power law, facilitates the observation of the critical behavior in LaPt$_2$Si$_2$. Also in this case, the inelastic part of the $q$ and energy scans were fitted to Lorentzian lineshapes while the elastic part was modeled with Gaussian lineshapes [Fig. \ref{ord_par_twin}(a-c)]. Figure \ref{ord_par_twin}(d-e) shows the raw energy scans normalized to the monitor counts at different $q$ points across the critical region. From these plots it is possible to follow both the $T-$dependence [Fig.~\ref{ord_par_twin}(f)] as well as the $q-$dependence of the soft phonon mode at the highest accessible temperature 470 K and below the CDW transition 210 K [Fig.~\ref{ord_par_twin}(g)]. The mode undergoes a remarkable $q$-dependent damping already at 470 K, with a minimum in the value of the phonon frequency at qk = 0.45 (r.l.u.). At $T=210$~K the phonon shows a $q$-dependent damping and shifting in frequency until partial collapse on the elastic line.
The fit of the temperature dependence of the phonon frequency to the power law $E_{tran} \propto (\frac{T}{T_c} - 1)^{\beta}$ [Fig.~\ref{ord_par_twin}(f)] provides a critical exponent equal to $\beta = 0.28 \pm 0.03$, which implies a phase transition with a non conventional critical behavior \cite{brusdeylins1989determination}. This fit provides also a transition temperature to the CDW state equal to $T_{CDW}$ = (237 $\pm$ 4)K.

\bibliography{sample}

\begin{thebibliography}{10}
\urlstyle{rm}
\expandafter\ifx\csname url\endcsname\relax
  \def\url#1{\texttt{#1}}\fi
\expandafter\ifx\csname urlprefix\endcsname\relax\def\urlprefix{URL }\fi
\expandafter\ifx\csname doiprefix\endcsname\relax\def\doiprefix{DOI: }\fi
\providecommand{\bibinfo}[2]{#2}
\providecommand{\eprint}[2][]{\url{#2}}

\bibitem{gruner1988dynamics}
\bibinfo{author}{Gr{\"u}ner, G.}
\newblock \bibinfo{journal}{\bibinfo{title}{The dynamics of charge-density
  waves}}.
\newblock {\emph{\JournalTitle{Reviews of Modern Physics}}}
  \textbf{\bibinfo{volume}{60}}, \bibinfo{pages}{1129} (\bibinfo{year}{1988}).

\bibitem{frohlich1954theory}
\bibinfo{author}{Fr{\"o}hlich, H.}
\newblock \bibinfo{journal}{\bibinfo{title}{On the theory of superconductivity:
  The one-dimensional case}}.
\newblock {\emph{\JournalTitle{Proceedings of the Royal Society of London.
  Series A. Mathematical and Physical Sciences}}}
  \textbf{\bibinfo{volume}{223}}, \bibinfo{pages}{296--305}
  (\bibinfo{year}{1954}).

\bibitem{da2014ubiquitous}
\bibinfo{author}{da~Silva~Neto, E.~H.} \emph{et~al.}
\newblock \bibinfo{journal}{\bibinfo{title}{Ubiquitous interplay between charge
  ordering and high-temperature superconductivity in cuprates}}.
\newblock {\emph{\JournalTitle{Science}}} \textbf{\bibinfo{volume}{343}},
  \bibinfo{pages}{393--396} (\bibinfo{year}{2014}).

\bibitem{wagner2008tuning}
\bibinfo{author}{Wagner, K.} \emph{et~al.}
\newblock \bibinfo{journal}{\bibinfo{title}{Tuning the charge density wave and
  superconductivity in {CuxTaS2}}}.
\newblock {\emph{\JournalTitle{Physical Review B}}}
  \textbf{\bibinfo{volume}{78}}, \bibinfo{pages}{104520}
  (\bibinfo{year}{2008}).

\bibitem{wang2013crossover}
\bibinfo{author}{Wang, A.} \emph{et~al.}
\newblock \bibinfo{journal}{\bibinfo{title}{A crossover in the phase diagram of
  {NaFe1- xCoxAs} determined by electronic transport measurements}}.
\newblock {\emph{\JournalTitle{New Journal of Physics}}}
  \textbf{\bibinfo{volume}{15}}, \bibinfo{pages}{043048}
  (\bibinfo{year}{2013}).

\bibitem{kim2015mechanism}
\bibinfo{author}{Kim, S.}, \bibinfo{author}{Kim, K.} \& \bibinfo{author}{Min,
  B.}
\newblock \bibinfo{journal}{\bibinfo{title}{The mechanism of charge density
  wave in {Pt-based layered superconductors: SrPt2As2 and LaPt2Si2}}}.
\newblock {\emph{\JournalTitle{Scientific reports}}}
  \textbf{\bibinfo{volume}{5}}, \bibinfo{pages}{1--10} (\bibinfo{year}{2015}).

\bibitem{shen2020evolution}
\bibinfo{author}{Shen, B.} \emph{et~al.}
\newblock \bibinfo{journal}{\bibinfo{title}{Evolution of charge density wave
  order and superconductivity under pressure in {LaPt2Si2}}}.
\newblock {\emph{\JournalTitle{Physical Review B}}}
  \textbf{\bibinfo{volume}{101}}, \bibinfo{pages}{144501}
  (\bibinfo{year}{2020}).

\bibitem{NocerinoXRD}
\bibinfo{author}{Nocerino, E.} \emph{et~al.}
\newblock \bibinfo{journal}{\bibinfo{title}{Structural evolution and onset of
  the density wave transition in the {CDW} superconductor {LaPt2Si2} clarified
  with synchrotron {XRD}}}.
\newblock {\emph{\JournalTitle{arXiv:2211.12617}}}  (\bibinfo{year}{2022}).

\bibitem{hase2013electronic}
\bibinfo{author}{Hase, I.} \& \bibinfo{author}{Yanagisawa, T.}
\newblock \bibinfo{journal}{\bibinfo{title}{Electronic structure of
  {LaPt2Si2}}}.
\newblock {\emph{\JournalTitle{Physica C: Superconductivity}}}
  \textbf{\bibinfo{volume}{484}}, \bibinfo{pages}{59--61}
  (\bibinfo{year}{2013}).

\bibitem{gupta2016coexistence}
\bibinfo{author}{Gupta, R.}, \bibinfo{author}{Paramanik, U.},
  \bibinfo{author}{Ramakrishnan, S.}, \bibinfo{author}{Rajeev, K.} \&
  \bibinfo{author}{Hossain, Z.}
\newblock \bibinfo{journal}{\bibinfo{title}{Coexistence of superconductivity
  and a charge density wave in {LaPt2(Si1- xGex)2} (0 $\leq$ x $\leq$ 0.5)}}.
\newblock {\emph{\JournalTitle{Journal of Physics: Condensed Matter}}}
  \textbf{\bibinfo{volume}{28}}, \bibinfo{pages}{195702}
  (\bibinfo{year}{2016}).

\bibitem{kubo2014structural}
\bibinfo{author}{Kubo, T.} \emph{et~al.}
\newblock \bibinfo{title}{Structural phase transition and superconductivity in
  {LaPt2Si2: 139La-and 195Pt-NMR} studies}.
\newblock In \emph{\bibinfo{booktitle}{Proceedings of the International
  Conference on Strongly Correlated Electron Systems (SCES2013)}},
  \bibinfo{pages}{017031} (\bibinfo{year}{2014}).

\bibitem{nagano2013charge}
\bibinfo{author}{Nagano, Y.} \emph{et~al.}
\newblock \bibinfo{journal}{\bibinfo{title}{Charge density wave and
  superconductivity of {RPt2Si2 (R= Y, La, Nd, and Lu)}}}.
\newblock {\emph{\JournalTitle{Journal of the Physical Society of Japan}}}
  \textbf{\bibinfo{volume}{82}}, \bibinfo{pages}{064715}
  (\bibinfo{year}{2013}).

\bibitem{falkowski2019structural}
\bibinfo{author}{Falkowski, M.}, \bibinfo{author}{Dole{\v{z}}al, P.},
  \bibinfo{author}{Andreev, A.}, \bibinfo{author}{Duverger-N{\'e}dellec, E.} \&
  \bibinfo{author}{Havela, L.}
\newblock \bibinfo{journal}{\bibinfo{title}{Structural, thermodynamic, thermal,
  and electron transport properties of single-crystalline {LaPt2Si2}}}.
\newblock {\emph{\JournalTitle{Physical Review B}}}
  \textbf{\bibinfo{volume}{100}}, \bibinfo{pages}{064103}
  (\bibinfo{year}{2019}).

\bibitem{aoyama2018195pt}
\bibinfo{author}{Aoyama, T.} \emph{et~al.}
\newblock \bibinfo{journal}{\bibinfo{title}{{195Pt-NMR} evidence for opening of
  partial charge-density-wave gap in layered {LaPt2Si2} with {CaBe2Ge2}
  structure}}.
\newblock {\emph{\JournalTitle{Journal of the Physical Society of Japan}}}
  \textbf{\bibinfo{volume}{87}}, \bibinfo{pages}{124713}
  (\bibinfo{year}{2018}).

\bibitem{falkowski2020multiple}
\bibinfo{author}{Falkowski, M.} \emph{et~al.}
\newblock \bibinfo{journal}{\bibinfo{title}{Multiple charge density wave states
  and magnetism in {NdPt2Si2} against the background of its nonmagnetic analog
  {LaPt2Si2}}}.
\newblock {\emph{\JournalTitle{Physical Review B}}}
  \textbf{\bibinfo{volume}{101}}, \bibinfo{pages}{174110}
  (\bibinfo{year}{2020}).

\bibitem{faak1997phonon}
\bibinfo{author}{F{\aa}k, B.} \& \bibinfo{author}{Dorner, B.}
\newblock \bibinfo{journal}{\bibinfo{title}{Phonon line shapes and excitation
  energies}}.
\newblock {\emph{\JournalTitle{Physica B: Condensed Matter}}}
  \textbf{\bibinfo{volume}{234}}, \bibinfo{pages}{1107--1108}
  (\bibinfo{year}{1997}).

\bibitem{stuhr2017thermal}
\bibinfo{author}{Stuhr, U.} \emph{et~al.}
\newblock \bibinfo{journal}{\bibinfo{title}{The thermal
  triple-axis-spectrometer {EIGER} at the continuous spallation source
  {SINQ}}}.
\newblock {\emph{\JournalTitle{Nuclear Instruments and Methods in Physics
  Research Section A: Accelerators, Spectrometers, Detectors and Associated
  Equipment}}} \textbf{\bibinfo{volume}{853}}, \bibinfo{pages}{16--19}
  (\bibinfo{year}{2017}).

\bibitem{kohn1959image}
\bibinfo{author}{Kohn, W.}
\newblock \bibinfo{journal}{\bibinfo{title}{Image of the fermi surface in the
  vibration spectrum of a metal}}.
\newblock {\emph{\JournalTitle{Physical Review Letters}}}
  \textbf{\bibinfo{volume}{2}}, \bibinfo{pages}{393} (\bibinfo{year}{1959}).

\bibitem{weber2011extended}
\bibinfo{author}{Weber, F.} \emph{et~al.}
\newblock \bibinfo{journal}{\bibinfo{title}{Extended phonon collapse and the
  origin of the charge-density wave in {2H-NbSe2}}}.
\newblock {\emph{\JournalTitle{Physical Review Letters}}}
  \textbf{\bibinfo{volume}{107}}, \bibinfo{pages}{107403}
  (\bibinfo{year}{2011}).

\bibitem{naito1982electrical}
\bibinfo{author}{Naito, M.} \& \bibinfo{author}{Tanaka, S.}
\newblock \bibinfo{journal}{\bibinfo{title}{Electrical transport properties in
  {2H-NbS2,-NbSe2,-TaS2 and-TaSe2}}}.
\newblock {\emph{\JournalTitle{Journal of the Physical Society of Japan}}}
  \textbf{\bibinfo{volume}{51}}, \bibinfo{pages}{219--227}
  (\bibinfo{year}{1982}).

\bibitem{borisenko2009two}
\bibinfo{author}{Borisenko, S.} \emph{et~al.}
\newblock \bibinfo{journal}{\bibinfo{title}{Two energy gaps and fermi-surface
  “arcs” in {NbSe2}}}.
\newblock {\emph{\JournalTitle{Physical review Letters}}}
  \textbf{\bibinfo{volume}{102}}, \bibinfo{pages}{166402}
  (\bibinfo{year}{2009}).

\bibitem{hodeau1978charge}
\bibinfo{author}{Hodeau, J.} \emph{et~al.}
\newblock \bibinfo{journal}{\bibinfo{title}{Charge-density waves in {NbSe3} at
  145k: Crystal structures, x-ray and electron diffraction studies}}.
\newblock {\emph{\JournalTitle{Journal of Physics C: Solid State Physics}}}
  \textbf{\bibinfo{volume}{11}}, \bibinfo{pages}{4117} (\bibinfo{year}{1978}).

\bibitem{monceau1976electric}
\bibinfo{author}{Monceau, P.}, \bibinfo{author}{Ong, N.},
  \bibinfo{author}{Portis, A.~M.}, \bibinfo{author}{Meerschaut, A.} \&
  \bibinfo{author}{Rouxel, J.}
\newblock \bibinfo{journal}{\bibinfo{title}{Electric field breakdown of
  charge-density-wave—induced anomalies in {NbSe3}}}.
\newblock {\emph{\JournalTitle{Physical Review Letters}}}
  \textbf{\bibinfo{volume}{37}}, \bibinfo{pages}{602} (\bibinfo{year}{1976}).

\bibitem{zhu2017misconceptions}
\bibinfo{author}{Zhu, X.}, \bibinfo{author}{Guo, J.}, \bibinfo{author}{Zhang,
  J.} \& \bibinfo{author}{Plummer, E.}
\newblock \bibinfo{journal}{\bibinfo{title}{Misconceptions associated with the
  origin of charge density waves}}.
\newblock {\emph{\JournalTitle{Advances in Physics: X}}}
  \textbf{\bibinfo{volume}{2}}, \bibinfo{pages}{622--640}
  (\bibinfo{year}{2017}).

\bibitem{yang2012charge}
\bibinfo{author}{Yang, J.} \emph{et~al.}
\newblock \bibinfo{journal}{\bibinfo{title}{Charge-orbital density wave and
  superconductivity in the strong spin-orbit coupled {IrTe2: Pd}}}.
\newblock {\emph{\JournalTitle{Physical review Letters}}}
  \textbf{\bibinfo{volume}{108}}, \bibinfo{pages}{116402}
  (\bibinfo{year}{2012}).

\bibitem{becker1999strongly}
\bibinfo{author}{Becker, B.} \emph{et~al.}
\newblock \bibinfo{journal}{\bibinfo{title}{Strongly coupled charge-density
  wave transition in single-crystal {Lu5Ir4Si10}}}.
\newblock {\emph{\JournalTitle{Physical Review B}}}
  \textbf{\bibinfo{volume}{59}}, \bibinfo{pages}{7266} (\bibinfo{year}{1999}).

\bibitem{moncton1975study}
\bibinfo{author}{Moncton, D.}, \bibinfo{author}{Axe, J.} \&
  \bibinfo{author}{DiSalvo, F.}
\newblock \bibinfo{journal}{\bibinfo{title}{Study of superlattice formation in
  {2H-NbSe2 and 2H-TaSe2} by neutron scattering}}.
\newblock {\emph{\JournalTitle{Physical Review Letters}}}
  \textbf{\bibinfo{volume}{34}}, \bibinfo{pages}{734} (\bibinfo{year}{1975}).

\bibitem{chen1981direct}
\bibinfo{author}{Chen, C.}, \bibinfo{author}{Gibson, J.} \&
  \bibinfo{author}{Fleming, R.}
\newblock \bibinfo{journal}{\bibinfo{title}{Direct observation of
  charge-density-wave discommensurations and dislocations in {2H-TaSe2}}}.
\newblock {\emph{\JournalTitle{Physical Review Letters}}}
  \textbf{\bibinfo{volume}{47}}, \bibinfo{pages}{723} (\bibinfo{year}{1981}).

\bibitem{mcmillan1976theory}
\bibinfo{author}{McMillan, W.~L.}
\newblock \bibinfo{journal}{\bibinfo{title}{Theory of discommensurations and
  the commensurate-incommensurate charge-density-wave phase transition}}.
\newblock {\emph{\JournalTitle{Physical Review B}}}
  \textbf{\bibinfo{volume}{14}}, \bibinfo{pages}{1496} (\bibinfo{year}{1976}).

\bibitem{brusdeylins1989determination}
\bibinfo{author}{Brusdeylins, G.} \emph{et~al.}
\newblock \bibinfo{journal}{\bibinfo{title}{Determination of the critical
  exponent for a charge density wave transition in {2H-TaSe2} by helium atom
  scattering}}.
\newblock {\emph{\JournalTitle{EPL (Europhysics Letters)}}}
  \textbf{\bibinfo{volume}{9}}, \bibinfo{pages}{563} (\bibinfo{year}{1989}).

\bibitem{weber2011electron}
\bibinfo{author}{Weber, F.} \emph{et~al.}
\newblock \bibinfo{journal}{\bibinfo{title}{Electron-phonon coupling and the
  soft phonon mode in {TiSe2}}}.
\newblock {\emph{\JournalTitle{Physical review Letters}}}
  \textbf{\bibinfo{volume}{107}}, \bibinfo{pages}{266401}
  (\bibinfo{year}{2011}).

\bibitem{das2018multigap}
\bibinfo{author}{Das, D.} \emph{et~al.}
\newblock \bibinfo{journal}{\bibinfo{title}{Multigap superconductivity in the
  charge density wave superconductor {LaPt2Si2}}}.
\newblock {\emph{\JournalTitle{Physical Review B}}}
  \textbf{\bibinfo{volume}{97}}, \bibinfo{pages}{184509}
  (\bibinfo{year}{2018}).

\bibitem{nie2021nodeless}
\bibinfo{author}{Nie, Z.} \emph{et~al.}
\newblock \bibinfo{journal}{\bibinfo{title}{Nodeless superconductivity in the
  charge density wave superconductor {LaPt2Si2}}}.
\newblock {\emph{\JournalTitle{Physical Review B}}}
  \textbf{\bibinfo{volume}{103}}, \bibinfo{pages}{014515}
  (\bibinfo{year}{2021}).

\bibitem{gold2018acoustic}
\bibinfo{author}{Gold-Parker, A.} \emph{et~al.}
\newblock \bibinfo{journal}{\bibinfo{title}{Acoustic phonon lifetimes limit
  thermal transport in methylammonium lead iodide}}.
\newblock {\emph{\JournalTitle{Proceedings of the National Academy of
  Sciences}}} \textbf{\bibinfo{volume}{115}}, \bibinfo{pages}{11905--11910}
  (\bibinfo{year}{2018}).

\bibitem{wei2021phonon}
\bibinfo{author}{Wei, B.}, \bibinfo{author}{Sun, Q.}, \bibinfo{author}{Li, C.}
  \& \bibinfo{author}{Hong, J.}
\newblock \bibinfo{journal}{\bibinfo{title}{Phonon anharmonicity: a pertinent
  review of recent progress and perspective}}.
\newblock {\emph{\JournalTitle{Science China Physics, Mechanics \& Astronomy}}}
  \textbf{\bibinfo{volume}{64}}, \bibinfo{pages}{1--34} (\bibinfo{year}{2021}).

\bibitem{voneshen2013suppression}
\bibinfo{author}{Voneshen, D.} \emph{et~al.}
\newblock \bibinfo{journal}{\bibinfo{title}{Suppression of thermal conductivity
  by rattling modes in thermoelectric sodium cobaltate}}.
\newblock {\emph{\JournalTitle{Nature materials}}}
  \textbf{\bibinfo{volume}{12}}, \bibinfo{pages}{1028--1032}
  (\bibinfo{year}{2013}).

\bibitem{voneshen2017hopping}
\bibinfo{author}{Voneshen, D.}, \bibinfo{author}{Walker, H.},
  \bibinfo{author}{Refson, K.} \& \bibinfo{author}{Goff, J.}
\newblock \bibinfo{journal}{\bibinfo{title}{Hopping time scales and the
  phonon-liquid electron-crystal picture in thermoelectric copper selenide}}.
\newblock {\emph{\JournalTitle{Physical Review Letters}}}
  \textbf{\bibinfo{volume}{118}}, \bibinfo{pages}{145901}
  (\bibinfo{year}{2017}).

\bibitem{wakamura1990observation}
\bibinfo{author}{Wakamura, K.}, \bibinfo{author}{Miura, F.},
  \bibinfo{author}{Kojima, A.} \& \bibinfo{author}{Kanashiro, T.}
\newblock \bibinfo{journal}{\bibinfo{title}{Observation of anomalously
  increasing phonon damping constant in the $\beta$ phase of the fast-ionic
  conductor {Ag 3 SI}}}.
\newblock {\emph{\JournalTitle{Physical Review B}}}
  \textbf{\bibinfo{volume}{41}}, \bibinfo{pages}{2758} (\bibinfo{year}{1990}).

\bibitem{gupta2018thermal}
\bibinfo{author}{Gupta, R.}, \bibinfo{author}{Rajeev, K.} \&
  \bibinfo{author}{Hossain, Z.}
\newblock \bibinfo{journal}{\bibinfo{title}{Thermal transport studies on charge
  density wave materials {LaPt2Si2 and PrPt2Si2}}}.
\newblock {\emph{\JournalTitle{Journal of Physics: Condensed Matter}}}
  \textbf{\bibinfo{volume}{30}}, \bibinfo{pages}{475603}
  (\bibinfo{year}{2018}).

\bibitem{bell2008cooling}
\bibinfo{author}{Bell, L.~E.}
\newblock \bibinfo{journal}{\bibinfo{title}{Cooling, heating, generating power,
  and recovering waste heat with thermoelectric systems}}.
\newblock {\emph{\JournalTitle{Science}}} \textbf{\bibinfo{volume}{321}},
  \bibinfo{pages}{1457--1461} (\bibinfo{year}{2008}).

\bibitem{momma}
\bibinfo{author}{Momma, K.} \& \bibinfo{author}{Izumi, F.}
\newblock \bibinfo{journal}{\bibinfo{title}{{VESTA}: a three-dimensional
  visualization system for electronic and structural analysis}}.
\newblock {\emph{\JournalTitle{Journal of Applied Crystallography}}}
  \textbf{\bibinfo{volume}{41}}, \bibinfo{pages}{653--658}
  (\bibinfo{year}{2008}).

\bibitem{azuah2009dave}
\bibinfo{author}{Azuah, R.~T.} \emph{et~al.}
\newblock \bibinfo{journal}{\bibinfo{title}{{DAVE}: a comprehensive software
  suite for the reduction, visualization, and analysis of low energy neutron
  spectroscopic data}}.
\newblock {\emph{\JournalTitle{Journal of research of the National Institute of
  Standards and Technology}}} \textbf{\bibinfo{volume}{114}},
  \bibinfo{pages}{341} (\bibinfo{year}{2009}).

\bibitem{igor}
\bibinfo{author}{Wave{M}etrics}.
\newblock \bibinfo{title}{{IGOR} {P}ro, scientific data analysis software}.

\end{thebibliography}

\section*{Acknowledgements}

The TAS-INS measurements were performed at the Swiss Spallation Neutron Source (SINQ), at the EIGER spectrometer at the Paul Scherrer Institute in Villigen, Switzerland (beamtimes proposals: 20211069 and 20212576). The TOF-INS measurements were performed at the Materials and Life Science Experimental Facility (MLF), at the HRC spectrometer at the Japan Proton Accelerator Research Complex in Tokai, Japan (beamtime proposal: 2019B0421). The authors wish to thank the staff of PSI and J-PARC for the valuable help in the neutron spectroscopy experiments. The authors also wish to thank Dr J. Lass and Prof Dr C. Niedermayer for the fruitful discussions and support during the EIGER experiments.
This research is funded by the Swedish Foundation for Strategic Research (SSF) within the Swedish national graduate school in neutron scattering (SwedNess). A.M. would like to acknowledge financial support from the E.R.C. (Grant 788144). Y.S. is funded by the Chalmers Area of Advance - Materials Science. 

\section*{Author contributions statement}

E.N. conceived the experiments. E.N., U.S., I.S.L., F.M., J.H., S.H., S.A., T.M. conducted the experiments. E.N., I.S., F.M. analyzed the results. The samples were synthesized by Z.H. and A.T. who also conducted the initial sample characterizations. E.N. and M.M. made all the figures. E.N. created the first draft, and all co-authors reviewed and revised the manuscript. 

\section*{Data availability statement}

All the data of this work are available from the corresponding authors on request. The data are also stored in the repositories of PSI and available from the EIGER instrument responsibles on request. 

\textbf{Competing interests} 

The authors declare no competing interests.  

\end{document}